\documentclass[final,5p,times,twocolumn,authoryear]{elsarticle}
\usepackage{jss-submission}
\usepackage{fontawesome5}
\usepackage{xcolor}

\journal{Journal of Systems and Software}

\newcommand{\file}[1]{\texttt{#1}}
\newcommand{\reportnumber}{FERMILAB-PUB-26-0712-T}

\makeatletter
\let\ps@pprintTitle@orig\ps@pprintTitle
\def\ps@pprintTitle{%
  \ps@pprintTitle@orig
  \def\@oddhead{\hfill\normalfont\small\reportnumber}%
  \let\@evenhead\@oddhead
}
\makeatother

\title{Designing Collaborative AI-Driven Workflows for Scientific
  Software Engineering\tnoteref{submitted}}

\begin{document}
\begin{frontmatter}

\tnotetext[submitted]{Preprint}

\author[argonne]{Akash Dhruv}
\author[fnal]{Max Knobbe}
\author[fnal]{Pedro Machado}
\author[riken]{Anshu Dubey}

\affiliation[argonne]{%
  organization={Argonne National Laboratory},
  country={United States}
}

\affiliation[fnal]{%
  organization={Fermi National Accelerator Laboratory},
  country={United States}
}

\affiliation[riken]{%
  organization={RIKEN Center for Computational Science},
  country={Japan}
}

\begin{abstract}
Agentic artificial intelligence systems can carry out a broad range of tasks in software engineering and scientific research, from writing and translating code to running workflows for data analysis and visualization. In scientific computing, the difficulty is verifying that agent-generated code is both correct and understandable to teams whose members bring different areas of expertise. We therefore argue that these systems are best used within collaborative team structures rather than as full automation. In the workflows we propose, domain experts write the specification and plan, and agents operate under a deterministic orchestration pattern to write the target code. Each stage ends with a numerical comparison against the reference code and requires human review and approval before the next begins. We evaluate these workflows on the translation of a large high-energy physics application from Fortran to C++, running the same task under different orchestrators, design patterns, and models. Across fourteen experiments, a simple author--reviewer loop with enforced limits completed a comparable number of files to a multi-agent workflow at about one-third of the cost per file.
\end{abstract}


\end{frontmatter}

\section{Introduction}
\label{sec:intro}
Scientific software is a research instrument whose development is chronically under-resourced. 
A numerical code can usually only be modified safely by someone who understands the science it encodes, and such expertise is scarce. 
The codes are long-lived and carry decades of modifications and domain knowledge that is rarely documented in the published literature.
Previous work in research software engineering has argued that investment in software practice is an investment in science, and that long-lived applications need continuous maintenance to remain correct and usable~\citep{heroux2023basic,godoy2025qmcpack}. 
Any substantial change to a scientific codebase, such as adding features, refactoring, porting to new hardware, or translating it to another language, is therefore necessary but limited by available developer time.
Modernizing legacy code is a common example, where the aim is to preserve the encoded domain knowledge and keep the code usable~\citep{shipman2023evaluation}.

Large language models (LLMs) are now widely used as coding assistants~\citep{chen2021evaluating,jimenez2024swe}, and agentic tools place them inside loops in which they read files, run commands, and revise their work based on feedback~\citep{yang2024sweagent,liu2024agentbench}.
This can lower the cost of software development and maintenance, but an agent's output is not guaranteed to be correct and its behavior over long tasks is difficult to control~\citep{lahlou2026systematic}. 
Because agentic systems require human supervision, unconstrained agent behavior directly increases supervisor workload.
Moreover, an agent overseen by a single person can pass an existing regression suite while introducing errors the suite misses. 
Such errors accumulate and make the code harder to maintain, which poses a risk to the quality and numerical correctness of scientific software.

We argue that agents are better used within a collaborative structure than in full automation, with work divided into stages that humans can easily review and verify. 
In the workflows we propose, domain experts write the specification and plan for each stage of the work, and agents carry out the mechanical work in bounded orchestration patterns. 
Each stage ends with a numerical comparison against the reference code, and only a human can record the approval that lets the next stage proceed. 
Building on empirical studies of the effect of AI coding assistants on developer productivity and code quality~\citep{ziegler2024copilot,dakhel2022copilot}, we explicitly separate the roles of supervision, implementation, and acceptance.
This keeps the human developer engaged and preserves their understanding of the code.

To test this approach, we use a computational Lab Notebook built on the practices we developed in our earlier work on reproducible computational research~\citep{dhruv2023provenance}, which provides a controlled environment in which the same task can be repeated under equivalent conditions.

The Lab Notebook is a versioned directory tree that holds the code, configuration, and the agentic layer (specification, plan, tools, and logs) for each stage. Keeping this state in shared files rather than in a user's private orchestrator configuration is what lets the same task be rerun under a different orchestrator for comparison.
As in the experimental sciences, the Lab Notebook provides consistent record keeping and reproducibility for computational work~\citep{graziani2021labmanager,oneal2023labnotebook}.

The test case we examine here is the translation of Monte Carlo for FeMtobarn processes (MCFM)~\citep{campbell1999update,campbell2011vector,campbell2019precision,campbell2022accelerating}, a widely used parton-level Monte Carlo program for precision collider phenomenology, written in Fortran. Modern high-performance computing (HPC) requirements make translation to C++ and a more modular software architecture desirable, with the eventual goal of integrating it into device-portable codes such as Pepper~\citep{Bothmann:2023gew}.
Code translation is a suitable test case because the existing code provides a numerical reference for each translated routine, so correctness can be measured directly. Translation is also already one of the main applications of LLMs in scientific computing and an active research area~\citep{assuncao2025modernization,pietrini2024bridging,valero2025fortrancpp,dearing2024lassi,davis2025pareval}.
In our earlier work, we used CodeScribe\footnote{CodeScribe is an open-source orchestrator written in Python, targeting scientific computing use cases, released under the Apache 2.0 license: \url{https://github.com/Lab-Notebooks/CodeScribe}.}, an LLM-driven tool, to generate Fortran--C++ interface layers for MCFM, allowing existing and translated code to run together during an incremental migration~\citep{dhruv2025leveraging}. In this paper, we extend CodeScribe's architecture to design autonomous agentic workflows and evaluate their performance and efficiency.

We run the same translation task, starting from the same commit and the same agentic layer, under different strategies and models. The strategies range from a simple author--reviewer loop to a multi-agent workflow running concurrent agents. In our experiments, the control needed to use agents on scientific software came from deliberate design of the workflow architecture and the agent execution policy. Adding concurrency without careful design did not improve the outcome and increased the cost.

The paper is organized as follows. Section~\ref{sec:architecture} states five design principles and describes the workflow architectures we examine. Section~\ref{sec:casestudy} presents the case study and the experimental design. Section~\ref{sec:evaluation} reports the evaluation and results, and Section~\ref{sec:conclusion} concludes with the implications and limitations.

\section{Collaborative Workflow Architecture}
\label{sec:architecture}
Domain experts and software engineers write the agentic specification, plan, and tools. An orchestrator uses them to drive subagents that produce code, and the code returns to the domain experts for acceptance at the verification boundary.

\subsection{Principles}
\label{sec:principles}

\textbf{P1. Verification at every stage boundary.}
A stage of work is finished when a numerical check accepted by the domain experts has passed. Successful compilation and the agent's own report are not sufficient for acceptance.

\textbf{P2. Humans design and accept the objectives, agents do the repetitive work.}
Agents are used for tasks that are broad and repetitive, such as applying a known translation rule across many files, updating build files, and checking test coverage. Humans make the decisions that require judgment, such as what counts as correct for a given routine, how to resolve a numerical disagreement, and whether to accept a result.

\textbf{P3. Reproducibility and provenance.}
The orchestration of the agents is deterministic, but their output is not. The workflow therefore records what was done, including which files were produced, which test exercised them, and what numerical agreement was observed. This record is kept in plain text files, tool-call logs, and reasoning summaries, so that a run can be inspected, resumed, or repeated.

\textbf{P4. Collaboration between roles.}
The work involves domain experts, who know what the code computes, and research software engineers, who know how it is built and run. Domain experts write the specification and review the results, while software engineers write and maintain the tools and workflow code that the agents use.

\textbf{P5. Open and user-controlled agent execution policy.}
The scientists doing the work should have full access to the orchestrator's agentic stack: system prompts, inter-agent interactions, tool definitions and protocols, model calls, shell environment behavior, and ReAct-based execution policies \citep{yao2023reactsynergizingreasoningacting}. With this control, workflows can be specified explicitly and reproduced reliably.

\begin{figure*}
  \centering
  \resizebox{0.82\textwidth}{!}{%
  \begin{tikzpicture}[
    font=\small,
    node distance=1.2cm,
    >={Latex[length=1.8mm, width=1.4mm]},
    every path/.style={line width=0.6pt},
    base/.style={draw, line width=0.6pt, rounded corners=1pt, align=center,
      inner sep=4pt},
    role/.style={base, line width=0.9pt, fill=white,
      minimum width=2.6cm, minimum height=0.95cm},
    file/.style={base, fill=gray!10,
      minimum width=2.9cm, minimum height=0.75cm},
    proc/.style={base, fill=white,
      minimum width=1.9cm, minimum height=0.75cm},
    vbar/.style={base, line width=0.9pt, fill=gray!25,
      minimum width=2.0cm, minimum height=0.95cm},
    grp/.style={font=\footnotesize\scshape, text=black!60},
  ]
    \node[role] (sci) {Domain\\scientists};
    \node[role, below=1.5cm of sci] (rse) {Research software\\engineers};

   \node[file, right=1.2cm of sci, yshift=0.3cm] (spec)
      {Spec\\[-1pt]{\footnotesize\itshape\color{black!70} requirements \& tests}};
    \node[file, below=0.3cm of spec] (plan)
      {Plan\\[-1pt]{\footnotesize\itshape\color{black!70} task-specific instructions}};
    \node[file, below=0.3cm of plan] (skill)
      {Tools\\[-1pt]{\footnotesize\itshape\color{black!70} and workflow code}};

    \node[proc, right=1.2cm of plan] (engine) {Orchestrator};
    \node[proc, right=0.7cm of engine] (sub) {Subagents};
    \node[proc, fill=gray!10, right=0.7cm of sub] (code) {Generated\\code};
    \node[vbar, right=0.7cm of code] (vb) {Verification\\boundary};

    \draw[->] (sci.east) -- ([yshift=4pt]spec.west);
    \draw[->] (sci.east) -- ([yshift=4pt]plan.west);
    \draw[->] (rse.east) -- ([yshift=-4pt]spec.west);
    \draw[->] (rse.east) -- ([yshift=-4pt]plan.west);
    \draw[->] (rse.east) -- (skill.west);

    \draw[->] (spec.east)  -- ([yshift=5pt]engine.west);
    \draw[->] (plan.east)  -- (engine.west);
    \draw[->] (skill.east) -- ([yshift=-5pt]engine.west);

    \draw[->] (engine) -- (sub);
    \draw[->] (sub)    -- (code);
    \draw[->] (code)   -- (vb);

    \coordinate (acctop) at ($(sci.north)+(0,0.9)$);
    \draw[->, dashed, dash pattern=on 3pt off 2pt] (vb.north) |- (acctop)
      node[pos=0.62, above, font=\footnotesize\itshape] {acceptance}
      -- (sci.north);

    \path (rse.south) ++(0,-0.3) coordinate (ly);
    \coordinate (exec) at ($(engine)!0.5!(code)$);
    \node[grp, anchor=north] at (sci  |- ly) {Authors};
    \node[grp, anchor=north] at (plan |- ly) {Shared artifacts};
    \node[grp, anchor=north] at (exec |- ly) {Agent execution};
  \end{tikzpicture}%
  } 
  \caption{Authoring responsibilities and the flow from human-authored inputs to accepted code. See Section~\ref{sec:architecture} for details.}
  \label{fig:provenance}
\end{figure*}
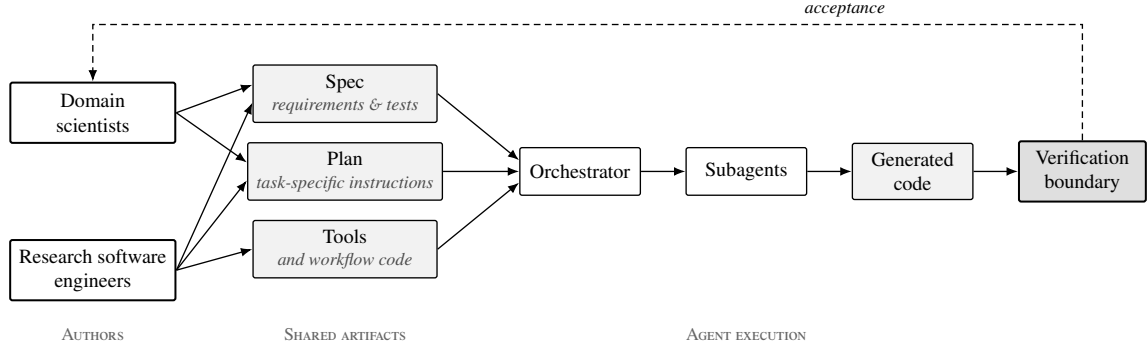

\subsection{Lab Notebook}
\label{sec:layout}

The principles above require a record kept outside any single agent or orchestrator: a place where the domain experts write the agentic layer and record approvals, which the agents can read but not modify (P2, P5), which retains what each run did (P3), and which is available to every member of the team and to every orchestrator (P4).
The Lab Notebook provides this record as a single versioned directory tree shared by the team.
\begin{figure}
  \centering
  \resizebox{0.4\textwidth}{!}{%
  \begin{tikzpicture}[>={Stealth[round]}, every node/.style={font=\footnotesize}]
    \node[draw=black!25, rounded corners=3pt, inner sep=7pt, fill=black!2]
      (tree) {%
      \begin{forest} dir tree
        [lab-notebook/
           [.claude/, for tree={fill=blue!9}
             [workflows/$\langle$workflow$\rangle$.js]
           ]
          [environment.sh]
          [sites/$\langle$user$\rangle$/config.sh]
          [software/
            [pepper/]
            [mcfm/]
          ]
          [tests/
            [pepper/]
            [mcfm/]
          ]
          [dev/, for tree={fill=green!10}
            [tools/
              [$\langle$name$\rangle$/]
            ]
            [transformations/
              [$\langle$name$\rangle$/
                  [desired\_spec.md]
                  [current\_plan.md]
                  [agent\_log.md]
                  [approvals.toml]
                  [loop.toml]
              ]
            ]
          ]
        ]
      \end{forest}%
    };
    \coordinate (aTop) at ([xshift=12mm,yshift=17mm]tree.north west);
    \coordinate (bTop) at ([xshift=-12mm,yshift=17mm]tree.north east);
    \path (aTop) pic{teammate={black}};
    \path (bTop) pic{teammate={black}};
    \node[above=1.5pt of aTop, font=\ttfamily\scriptsize, black] 
    {collaborator 1};
    \node[above=1.5pt of bTop, font=\ttfamily\scriptsize, black]
      {collaborator 2};

    \draw[<->, semithick, black]
      ([yshift=-14pt]aTop) -- (aTop |- tree.north);
    \draw[<->, semithick, black]
      ([yshift=-14pt]bTop) -- (bTop |- tree.north);

    \node[below=3mm of tree.south, font=\scriptsize, align=left,
          draw=black!20, rounded corners=2pt, fill=black!2, inner sep=4pt]
      (legend) {%
      \tikz[baseline={([yshift=-0.5ex]current bounding box.center)}]{%
        \node[fill=green!10, draw=black!25, rounded corners=1pt,
          minimum width=3.4mm, minimum height=2.6mm, inner sep=0pt]{};}\;%
      \textbf{{dev/}}. Interaction layer for human-AI collaboration to modify source code};
  \end{tikzpicture}
  }
  \caption{Structure of the lab notebook for the MCFM case study. See Section~\ref{sec:layout} for details.}
  \label{fig:labnotebook}
\end{figure}
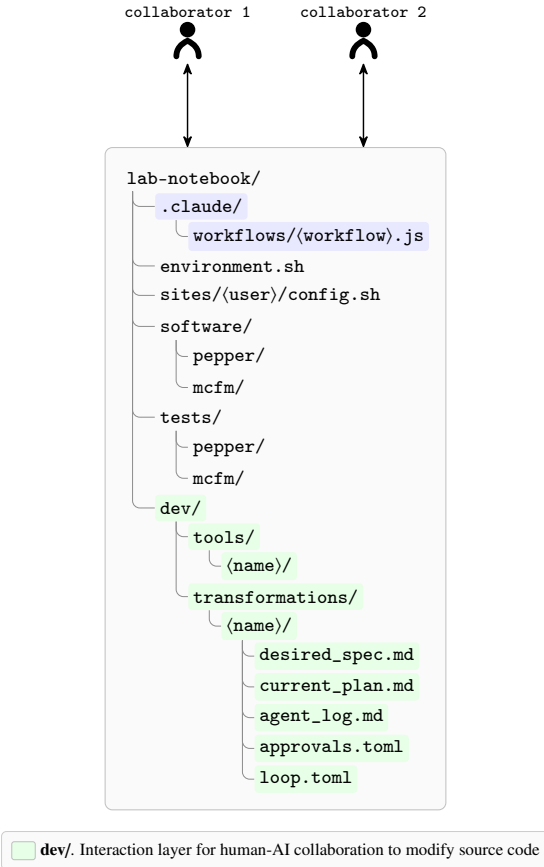

Figure~\ref{fig:labnotebook} shows how the tree is organized for our case study, with configuration inherited downward from the top of the project. 
The \file{environment.sh} file at the project root holds a common Unix shell configuration, and each collaborator's machine-specific settings are kept in their own subtree under \file{sites/$\langle$user$\rangle$/config.sh}, so each can adapt the environment to their system without changing the shared file. Each subdirectory of \file{software/} tracks an upstream GitHub repository, here the Pepper event generator and MCFM, and the matching directories under \file{tests/} follow each code's own build and test practices.

The directory \file{dev/} holds the files (agentic layer) through which humans and agents work on the code, namely the transformations and the tools described below. Some orchestrators also expect their own files at a fixed location. Claude Code, for example, reads workflow scripts from \file{.claude/workflows} at the root of the tree.

\begin{figure*}[t]
  \centering
  \resizebox{0.95\textwidth}{!}{
  \begin{tikzpicture}[
    font=\footnotesize,
    >={Latex[length=1.8mm, width=1.4mm]},
    every path/.style={line width=0.6pt},
    base/.style={draw, line width=0.6pt, rounded corners=1pt, align=center,
      inner sep=3pt},
    file/.style={base, fill=gray!10, minimum width=2.0cm,
      minimum height=0.38cm, font=\scriptsize},
    pattern/.style={base, line width=0.9pt, fill=gray!10, text width=2.1cm,
      minimum height=0.8cm},
    proc/.style={base, fill=white, text width=2.1cm, minimum height=0.52cm},
    small/.style={proc, text width=1.9cm},
    tier/.style={fill=gray!40, line width=0.9pt},
    lbl/.style={font=\scriptsize, text=black!60, align=center},
  ]
    \node[file] (spec) at (1.22,1.00) {Spec};
    \node[file] (plan) at (1.22,0.52) {Plan};
    \node[file] (tool) at (1.22,0.04) {Tools};
    \node[file] (logs) at (1.22,-0.44) {Session logs};
    \node[file] (code) at (1.22,-0.92) {Current code};
    \node[draw, dashed, line width=0.6pt, rounded corners=2pt,
      fit=(spec)(plan)(tool)(logs)(code), inner sep=0.22cm,
      label={[lbl]above:Shared inputs (on disk)}] (inputs) {};

    \node[pattern] (wf) at (4.50,1.09) {Multi-agent\\workflow};
    \node[proc] (disp)  at (7.61,1.09) {Dispatch\\[-1pt]{\scriptsize triage $+$ bundle}};
    \node[small] (a1)   at (10.67,1.78) {Author};
    \node[small] (a2)   at (10.67,1.24) {Author};
    \node[text=black!55, font=\large, inner sep=0pt] (dots) at (10.67,0.79) {$\vdots$};
    \node[small, tier] (a3) at (10.67,0.34) {Author};
    \node[proc, tier] (int) at (13.73,1.09) {Integrate\\[-1pt]{\scriptsize build $+$ verify}};
    \node[small, tier] (fix) at (17.10,1.09) {Fix};
    \node[lbl] at (10.67,2.32) {parallel};

    \node[pattern] (lp) at (4.50,-1.05) {Author--reviewer\\loop};
    \node[proc] (auth)  at (8.20,-1.05) {Author};
    \node[proc] (rev)   at (12.60,-1.05) {Reviewer};

    \draw[->] (inputs.east) -- (wf.west);
    \draw[->] (inputs.east) -- (lp.west);

    \draw[->] (wf) -- (disp);
    \draw[->] (disp.east) -- (a1.west);
    \draw[->] (disp.east) -- (a2.west);
    \draw[->] (disp.east) -- (a3.west);
    \draw[->] (a1.east) -- ([yshift=5pt]int.west);
    \draw[->] (a2.east) -- (int.west);
    \draw[->] (a3.east) -- ([yshift=-5pt]int.west);
    \draw[->] ([yshift=5pt]int.east)  -- node[lbl, above] {failed}
      ([yshift=5pt]fix.west);
    \draw[->] ([yshift=-5pt]fix.west) -- node[lbl, below] {re-integrate}
      ([yshift=-5pt]int.east);

    \draw[->] (lp) -- (auth);
    \draw[->] ([yshift=4pt]auth.east) -- node[lbl, above] {result}
      ([yshift=4pt]rev.west);
    \draw[->] ([yshift=-4pt]rev.west) -- node[lbl, below] {findings}
      ([yshift=-4pt]auth.east);
  \end{tikzpicture}
  }
  \caption{One set of human-authored inputs, two orchestrators: a multi-agent
    workflow (top) and an author--reviewer loop (bottom). Unshaded nodes are
    sessions that run under the model the workflow was launched with, and
    shaded nodes are pinned to a higher tier model (Opus~5 in our
    comparisons). The vertical dots stand for further parallel authors.
    Section~\ref{sec:stacks} gives the details.}
  \label{fig:inputs}
\end{figure*}
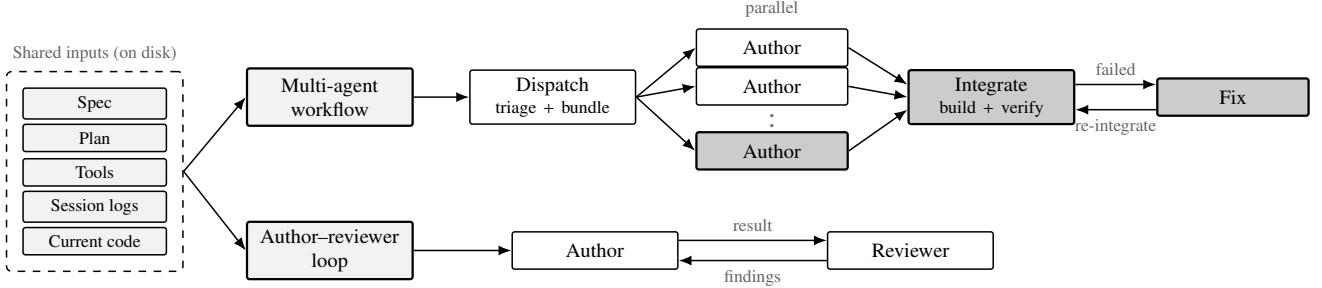

The agentic work changes the code from one state to another, for example from Fortran to C++, and a project may carry several of these transformations at once.
Each transformation is a directory under \file{dev/transformations} that contains a specification (\file{desired\_spec.md}), a plan (\file{current\_plan.md}), the agent log (\file{agent\_log.md}), and the configuration files through which humans interact with the agents (\file{approvals.toml}, \file{loop.toml}). 
Because this state is kept on disk, a transformation can be paused, resumed, or handed to a different orchestrator or a teammate.

The tools available to an agent are the internal tools of an orchestrator, programs already available in the shell environment, and project-specific scripts under \file{dev/tools}, which operate on a file or a list of files. The latter can be a collection of scripts, tied together with a common command-line interface like \file{dev/workflow.py} in the example Lab Notebook \citep{jss_example_2026}.

Humans have write access to every file in a transformation directory, and the agent only to its log.
The specification and the plan are written by humans and carry the intent and the acceptance criteria, and the approvals in \file{approvals.toml} are likewise read-only for the agents.
In \file{agent\_log.md} the agents record what was attempted, which checks were run, and what remains open.

The specification and plan files are written in Markdown.
Their content depends on the task, but across the transformations in the reference notebook \citep{jss_example_2026} they share a common structure, shown below with the task-specific text omitted. 
The plan in \file{current\_plan.md} states the steps the agent repeats in each round. In practice, a plan is similar to a skill, but it applies only to a specific transformation.
\begin{lstlisting}[style=mdstyle, title={current\_plan.md}]
# Plan

1. What to do each round.
2. How to write the agent log file.
3. How to check for human approvals.
4. How to use tools.
5. How to select units of work.
6. How to verify results.
7. When to stop.
\end{lstlisting}
The specification in \file{desired\_spec.md} states the contract of the transformation, the expected output, the type conversions and their caveats, and the correctness criterion.
\begin{lstlisting}[style=mdstyle, title={desired\_spec.md}]
# Spec

1. Contract.
2. Output shape.
3. Type conversions and their caveats.
4. Correctness bar.
\end{lstlisting}

\subsection{Staged work}
\label{sec:stages}

A large change to a scientific code is divided into stages separated by human review.
The stages are not fixed in advance. Developers choose them at each boundary, starting from a small part of the code that can be tested and expanding outward from it.
Within a stage the code is indexed, converted, and tested against the reference, and the result is then offered for review.
The stage boundaries are the verification points of P1 and the points where the roles of P4 meet.
Small stages can be checked against a concrete reference and assigned to team members by expertise.

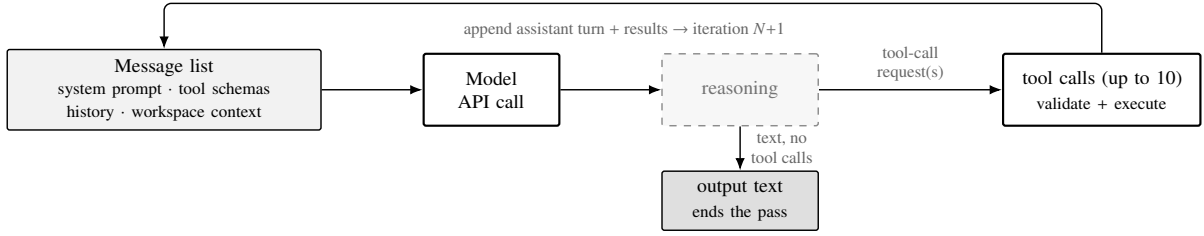
\begin{figure*}[t]
  \centering
  \resizebox{0.87\textwidth}{!}{
  \begin{tikzpicture}[
    font=\footnotesize,
    >={Latex[length=1.8mm, width=1.4mm]},
    every path/.style={line width=0.6pt},
    base/.style={draw, line width=0.6pt, rounded corners=1pt, align=center,
      inner sep=4pt, minimum height=1.05cm},
    state/.style={base, fill=gray!10, text width=4.3cm},
    model/.style={base, line width=0.9pt, fill=white, text width=1.7cm},
    reason/.style={base, draw=black!45, dashed, fill=black!3, text=black!55,
      text width=2.0cm},
    act/.style={base, line width=0.9pt, fill=white, text width=2.6cm},
    term/.style={base, fill=gray!25, text width=2.0cm, minimum height=0.8cm},
    lbl/.style={font=\scriptsize, text=black!60, align=center},
  ]
    \node[state] (msg) at (0,0) {Message list\\[1pt]{\scriptsize system prompt
      $\cdot$ tool schemas\\ history $\cdot$ workspace context}};
    \node[model] (api)    at (4.78,0)  {Model\\API call};
    \node[reason] (reason) at (8.41,0)  {reasoning};
    \node[act] (act)      at (13.69,0) {tool calls (up to 10)\\[1pt]{\scriptsize
      validate $+$ execute}};
    \node[term] (final)   at (8.41,-1.62) {output text\\[1pt]{\scriptsize ends
      the pass}};

    \draw[->] (msg) -- (api);
    \draw[->] (api) -- (reason);
    \draw[->] (reason.east) -- node[lbl, above] {tool-call\\request(s)}
      (act.west);
    \draw[->] (reason.south) -- node[lbl, right, xshift=1pt] {text, no\\tool
      calls} (final.north);

    \draw[->, rounded corners=3pt] (act.north) -- ++(0,0.72) -| (msg.north);
    \node[lbl] at (6.72,0.87) {append assistant turn $+$ results
      $\rightarrow$ iteration $N{+}1$};
  \end{tikzpicture}
  }
  \caption{One chat-completion iteration $N$ of a single agent, the primitive
    inside both the CodeScribe author and reviewer. Reasoning is shown in the
    lighter shade, and up to ten tool calls execute per iteration.
    Section~\ref{sec:stacks} describes the iteration mechanics. Budget
    exhaustion, which also ends the pass, is not drawn.}
  \label{fig:iteration}
\end{figure*}
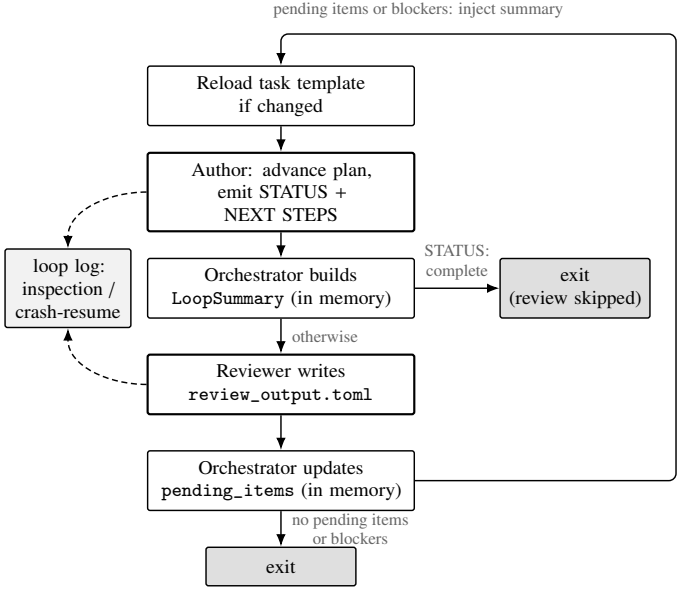
\begin{figure}
  \centering
  \resizebox{\columnwidth}{!}{%
  \begin{tikzpicture}[
    font=\footnotesize,
    >={Latex[length=1.8mm, width=1.4mm]},
    every path/.style={line width=0.6pt},
    base/.style={draw, line width=0.6pt, rounded corners=1pt, align=center,
      inner sep=4pt, text width=3.6cm, minimum height=0.62cm},
    proc/.style={base, fill=white},
    agent/.style={base, line width=0.9pt, fill=white},
    term/.style={base, fill=gray!25, text width=1.9cm, minimum height=0.56cm},
    art/.style={base, fill=gray!10, text width=1.55cm},
    lbl/.style={font=\scriptsize, text=black!60, align=center},
  ]
    \node[proc]  (reload) at (0,0)    {Reload task template\\if changed};
    \node[agent] (auth)   at (0,-1.4) {Author: advance plan,\\emit \mbox{STATUS}
      $+$ \mbox{NEXT STEPS}};
    \node[proc]  (summ)   at (0,-2.8) {Orchestrator builds\\\file{LoopSummary}
      (in memory)};
    \node[agent] (rev)    at (0,-4.2) {Reviewer writes\\\file{review\_output.toml}};
    \node[proc]  (upd)    at (0,-5.6) {Orchestrator updates\\\file{pending\_items}
      (in memory)};

    \node[term] (exit1) at (4.28,-2.8)  {exit\\(review skipped)};
    \node[term] (exit2) at (0,-6.85)    {exit};
    \node[art]  (log)   at (-3.10,-2.8) {loop log:\\inspection /\\crash-resume};

    \draw[->] (reload) -- (auth);
    \draw[->] (auth) -- (summ);
    \draw[->] (summ) -- node[lbl, right, xshift=1pt] {otherwise} (rev);
    \draw[->] (rev) -- (upd);
    \draw[->] (summ.east) -- node[lbl, above] {\mbox{STATUS:}\\complete}
      (exit1.west);
    \draw[->] (upd) -- node[lbl, right, xshift=1pt] {no pending items\\or blockers}
      (exit2);

    \draw[->, rounded corners=3pt] (upd.east) -- (5.75,-5.6) -- (5.75,0.90)
      -| (reload.north);
    \node[lbl] at (2.00,1.25) {pending items or blockers: inject summary};

    \draw[->, dashed, dash pattern=on 2.5pt off 1.5pt] (auth.west)
      to[out=180, in=90] (log.north);
    \draw[->, dashed, dash pattern=on 2.5pt off 1.5pt] (rev.west)
      to[out=180, in=-90] (log.south);
  \end{tikzpicture}%
  }
  \caption{One cycle of the CodeScribe author--reviewer loop. An author advances
    the Plan and a reviewer checks it, and a summary of what remains is carried
    into the next cycle (Section~\ref{sec:stacks}). Each agent runs its own
    multi-turn chat-completion iterations (Figure~\ref{fig:iteration}). The
    on-disk log files on the left are kept for inspection and crash-resume.}
  \label{fig:csloop}
\end{figure}
\subsection{Orchestration patterns}
\label{sec:stacks}

Because the inputs are kept separate from the orchestrator, the orchestration can be changed while the task and the correctness criterion stay fixed. 
We compare two patterns: an author--reviewer loop, and a phased workflow that distributes the work over many subagents and integrates their results serially (Figure~\ref{fig:inputs}).

Single-agent loops are a common choice for such translation tasks. One agent runs repeatedly in short sessions and keeps all of its state on disk, a pattern known as the Ralph Loop~\citep{huntley2025ralph} and used by \citet{lahlou2026systematic} for Fortran to JAX code translation.
Each iteration starts a new session and passes it a file, \file{prompt.md}, that points the agent to the shared inputs.
The session reads the current state, advances the work, and exits, so state passes between iterations only through the files on disk.
\citet{knuth2026claude} describes a similar practice, in which the agent records its progress in a plan file after every step.
The purpose is to avoid the degradation of output quality due to growing context length~\citep{liu2024lost, modarressi2025nolima, hong2025context}.

\textbf{CodeScribe author--reviewer loop.}
CodeScribe \citep{jss_codescribe_2026} implements a fixed author--reviewer loop that extends the single-agent loop. 
Two agents work on the shared inputs, an author that advances the plan and a reviewer that checks the result and passes its findings to the next cycle. 
The loop is implemented in \file{\_loop.py} in the CodeScribe repository and is configured by \file{loop.toml}. 
This file lists the shell commands that the author may run in addition to CodeScribe's internal tools (read, write, edit, glob).
It also contains a first user message, under the key \file{[[chat.user]]}, that CodeScribe places in the chat history before the loop starts.
The message points the agent to the specification and plan files, tells it to read the plan first, and states that the constraints it should operate under. The file serves the same role as \file{prompt.md} in a single-agent loop, while the instruction for each cycle is built by CodeScribe. The configuration used in the experiments is as follows.
\begin{lstlisting}[style=tomlstyle, title={loop.toml}]
[tools]
bash = ["jobrunner", "python3"]

[[chat.user]]
content = '''
Each round, follow current_plan.md and the contract in
desired_spec.md. Read the plan first, then the spec.

Authority: current_plan.md and desired_spec.md are
human-owned -- read, do not modify. agent_log.md is
the only file in dev/transformations/<name>/ that
the AI may modify. On conflict, the spec governs.
'''
\end{lstlisting}
The allowlist contains \file{jobrunner}~\citep{dhruv2023provenance}, which builds the code and runs the benchmarks, and \file{python3}, through which the agent calls \file{dev/workflow.py}.

The CodeScribe loop repeats at two levels. 
A cycle is one author pass followed by one review pass. 
Within a pass, the agent makes a sequence of model API calls, built on ReAct policies, which we refer to as iterations (Figure~\ref{fig:iteration}).
Each iteration sends the message list to the model and receives its reasoning, tool-call requests, and output text.
Between calls, CodeScribe validates each requested tool call, runs it only if it passes, and appends the result to the message list.
A pass ends when the model returns completion text without tool calls.

The CodeScribe loop, which runs at most five cycles by default, is shown in Figure~\ref{fig:csloop}. 
It contains only the agents and the orchestrator, and acceptance lies outside the loop.
At the end of each cycle, CodeScribe adds a summary of the cycle and the list of pending items to the prompt of the next cycle.
The loop ends earlier in two cases: when the author agent reports the work complete, which skips the review, or when the reviewer reports no pending items or blockers (pending items that require a human review).
CodeScribe also writes log files in \file{.codescribe/loop} for inspection, crash-resume, and audit.

\textbf{Enforced limits.}
CodeScribe enforces the limits on each agent at the validation step between iterations. It caps the number of iterations and tool calls, and it rejects a tool call after repeated failures.
After several consecutive failed iterations, it instructs the agent to report a blocker instead of continuing.
When a tool call is rejected, CodeScribe returns an error message to the agent so it can diagnose the problem in the next iteration.
Unlike limits stated in the specification or the prompt, which the model may not follow, these cannot be exceeded by the agent.
This implements principle P5 and allows the loop to run unsupervised within a stage.

\textbf{Implementation.}
CodeScribe calls the model APIs directly (e.g. Anthropic's Messages API or OpenAI's Chat Completions API). This gives it control over which tools the agents call and how much context each tool result adds.
CodeScribe is open source and takes configurable API keys, so the same tools and plan can be run against different providers, including open-weight models.
Working at the API level also gives direct control over prompt caching and reasoning. 
CodeScribe places explicit cache breakpoints on the system prompt, the tool schemas, and the accumulated history, so the cache written in one iteration is read in the next. 
The workspace context, which holds the iteration number, the remaining budgets, and recent tool results, is appended after the last breakpoint and rebuilt at every iteration, so it does not invalidate the cache. 
Reasoning is set per provider: on Anthropic's API it enables adaptive thinking with summarized traces, and on an OpenAI-compatible endpoint it sets a fixed high reasoning effort. 
The Anthropic models used here (Opus~5 and Sonnet~5) think adaptively by default, so for them the parameter does not change how much the model reasons, only whether a summary of the reasoning is returned and recorded in the logs. 
Section~\ref{sec:evaluation} shows an example of a CodeScribe transcript and reports API call measurements.

\textbf{Claude Code Workflows.}
\label{sec:orchestration}
A Claude Code workflow is a script under \file{.claude/workflows} that launches subagents. We wrote these scripts by prompting Claude, with minimal supervision for the multi-agent workflow and detailed supervision for the author--reviewer workflow, so that the latter matches the design of the CodeScribe loop and the comparison stays meaningful. The workflows run their agents in separate contexts (Figure~\ref{fig:inputs}), triggered by a natural language instruction:
{\small
\begin{verbatim}
Run /<workflow> for dev/transformations/<name>
\end{verbatim}
}

For the multi-agent workflow, authoring is parallel and integration is serial. Many subagents translate individual files at the same time, each writing only its own output, while a single integrator subagent owns the shared build, adds the new files, and runs the verification. Authoring is routed by file complexity: the workflow inspects each file and sends ordinary files to a lighter model and large or deeply nested files to a stronger one. To profile the full transformation end to end, we run the workflow as an open-ended process across these agents, so it terminates only at the approval gate rather than after a fixed number of steps. As a result, the multi-agent workflow itself runs as a loop, with additional control flow around the agents.

The author-reviewer loop workflow in Claude Code mirrors CodeScribe, but its limits on tool usage, shell environment, agent interactions, and ReAct policies are enforced only through prompt-based guidance, and it has no control over prompt caching or model API interactions. This workflow differs from the default \file{/loop} skill in Claude Code, and placing a \file{loop.js} in \file{.claude/workflows} overrides that default behavior.

Following our design principles, the agentic layer is tracked primarily in the Lab Notebook repository for crash-resume and reproducibility. However, Claude Code also keeps a user-specific memory directory outside the Lab Notebook, which can affect agent behavior. We discuss this in Section~\ref{sec:evaldecision}.

\section{Case Study: MCFM}
\label{sec:casestudy}
With the inputs fixed, the workflow configurations differ only in how the orchestrator is started and in how much control flow surrounds the agents. We investigate three consequences of that difference: whether the additional structure of a multi-agent workflow justifies its cost relative to the loop, whether enforcing limits in the orchestrator rather than stating them in the prompt changes what the agent does, and how far either result depends on the model.

We evaluate these effects on the modernization of MCFM, a parton-level Monte Carlo program for hadron-collider phenomenology. MCFM provides a large library of Standard Model processes. It computes cross sections and differential distributions at next-to-leading order (NLO) in QCD and, for a subset of processes, at next-to-next-to-leading order (NNLO). To do so, it also supplies the necessary phase-space integration, parton-distribution handling, and analysis routines~\citep{campbell2019precision}. The task is representative of a broader scientific software engineering problem: development and improvement of a Fortran codebase with hundreds of source files and deep call trees.

\subsection{Task, target and verification criteria}
\label{sec:task}
Translation is done in two steps.
The first step translates each Fortran unit into C++, emitting three
files per unit: a C++ implementation with an \file{extern "C"} wrapper,
a C++ header, and an \file{iso\_c\_binding} Fortran interface. 
These bindings let existing Fortran callers continue to link and run while the code is translated one unit at a time.  
A second step ports the resulting C++ code to
device-portable kernels, so the full pipeline is
Fortran~$\rightarrow$~C++~$\rightarrow$~kernels.

Correctness is defined by the code's own benchmarks. A translated file is verified when two checks pass. First, a coverage probe perturbs the translated file and confirms that the benchmark result changes, showing that the benchmark exercises it. Second, the unperturbed file reproduces the reference benchmark value within a fixed tolerance. Together these two checks form the oracle used in the rest of the paper. 
A file that no benchmark exercises, such as pure infrastructure code, is tagged as translated but not verified.

\subsection{The Lab Notebook and the Design of Experiments}
\label{sec:setup}\label{sec:doe}
The Lab Notebook's structure (Figure~\ref{fig:labnotebook}) is a template other teams can adopt for their own AI-driven workflows. The inheritance structure, the shared agentic layer design, and the staged transformation format are all generic. Only the tools under \file{dev/tools} are specific to MCFM. A different project can keep the same layout, change the project-specific parts, and add its own implementation of a concrete agentic layer with tools accessible via model context protocal (MCP).

The transformation \file{transformations/mcfm-translate} is designed to batch-translate a group of Fortran files.
Each file carries three attributes. A file is ready when every file it
depends on has been translated, and it is not marked as untestable
infrastructure. File-level dependencies are obtained from a Doxygen
cross-reference build of the Fortran source. Its oracle status is verified, translated, or failed, as determined by the
checks above. Its recorded outcome is the status the agent itself reports,
which may contain hallucinations.

The agent groups ready files, typically about five that share a folder or a test, and lists each group in \file{agent\_log.md} as a Markdown heading followed by a checklist of its files.
Before opening a new group, an agent runs a gate script that reads these groups and the approvals in \file{approvals.toml}. 
The gate limits how far the agents can work ahead of human review: a new group cannot be opened while three completed groups are waiting for approval, or while a group waiting for approval contains a failed file. 
Work in a group that is already open may finish. Only a human writes approvals to \file{approvals.toml} (Section~\ref{sec:layout}).

The deterministic steps, namely building the dependency map, ranking the
ready files, checking a file against the oracle, and enforcing the gate, are scripts under \file{dev/tools}. 
Under both orchestrators, the agents run these tools through \file{dev/workflow.py}, which takes a transformation name and runs the corresponding script using Python. CodeScribe agents use the \file{python3} executable from their allowlist, and Claude Code agents follow the directions in their prompts.
The gate and the oracle are therefore the same code in every run, so the orchestrator and the pattern can be varied while everything else stays fixed.

We ran \file{transformations/mcfm-translate} fourteen times. 
All runs start from the same
commit of the MCFM submodule and therefore from the same set of 445 candidate
files. Table~\ref{tab:coveragemap} lists the date of each run and the archived
directory that contains it~\citep{jss_example_2026}. 
We call each combination
of orchestrator and pattern a configuration, and test the following
three.

\begin{table*}[t]
  \centering
  \caption{Run manifest linking each of the fourteen runs to its exact
    directory in the archived lab notebook \citep{jss_example_2026}, the
    Zenodo snapshot of the repository that Figure~\ref{fig:labnotebook} illustrates.
    Paths are relative to the repository root.}
  \label{tab:coveragemap}
  \scriptsize
\setlength{\tabcolsep}{3.2pt}
\begin{tabular}{@{}lll@{}}
  \toprule
  Run & Configuration & Directory (under {\ttfamily evals/} in the archived repository) \\
  \midrule
    R1 & ccworkflow (sonnet-5 triage and dispatch, opus-5 integrate) & \texttt{experiments/08-27-2026/ccworkflow-sonnet-5-opus-5-integrate-run3} \\
    R2 & ccworkflow (opus-5 triage and dispatch) & \texttt{experiments/08-28-2026/ccworkflow-opus-5} \\
    R3 & ccworkflow (sonnet-5 triage and dispatch, opus-5 integrate, run2) & \texttt{experiments/08-28-2026/ccworkflow-sonnet-5-opus-5-integrate-run4} \\ \hline
    R4 & csloop opus-5 & \texttt{experiments/08-27-2026/codescribe-opus-5} \\
    R5 & csloop opus-5 (run2) & \texttt{experiments/08-27-2026/codescribe-opus-5-run2} \\
    R6 & csloop opus-5 (run3) & \texttt{experiments/08-28-2026/codescribe-opus-5} \\
    R7 & csloop sonnet-5 (run2) & \texttt{experiments/08-27-2026/codescribe-sonnet-5-run2} \\
    R8 & csloop sonnet-5 (run3, incomplete — no agent\_log archived) & \texttt{experiments/08-28-2026/codescribe-sonnet-5-run3} \\ \hline
    R9 & csloop gpt-5.6-sol (run4) & \texttt{experiments/08-27-2026/codescribe-oaic-gpt56sol-run4} \\
    R10 & csloop gpt-5.6-sol (run5) & \texttt{experiments/08-27-2026/codescribe-oaic-gpt56sol-run5} \\
    R11 & csloop gpt-5.6-sol (run6) & \texttt{experiments/08-28-2026/codescribe-oaic-gpt56sol-run6} \\ \hline
    R12 & ccloop opus-5 & \texttt{experiments/09-27-2026/ccworkflow-loop-opus-5-run3} \\
    R13 & ccloop opus-5 (run2) & \texttt{experiments/09-29-2026/ccworkflow-loop-opus-5-run5} \\
    R14 & ccloop sonnet-5 & \texttt{experiments/09-28-2026/ccworkflow-loop-sonnet-5-run3} \\
  \bottomrule
\end{tabular}

\end{table*}

\begin{table*}[t]
  \centering
  \caption{The fourteen \file{transformations/mcfm-translate} runs of
    Section~\ref{sec:doe}: three ccworkflow runs (R1--R3), eight csloop
    runs (R4--R11), and three ccloop runs (R12--R14). Files settled counts
    translations landed on each run's branch, including units whose Fortran
    original was left in place. \file{Loops} reports completed cycles against
    each run's configured cap and differs in kind by orchestrator
    (Section~\ref{sec:evalresults}). Figure~\ref{fig:evalcombined} shows
    the same data in normalized form.}
  \label{tab:runs}
  \footnotesize
\begin{tabular}{@{}llrrrrrrrr@{}}
  \toprule
  & & & \multicolumn{3}{c}{Run totals} & \multicolumn{3}{c}{Per file settled} & \\
  \cmidrule(lr){4-6}\cmidrule(lr){7-9}
  & Configuration & Loops & Files & Min & USD & USD & Min & Tool calls & Cache \%\\
  \midrule
    R1 & ccworkflow (sonnet-5 triage and dispatch, opus-5 integrate) & 3/3 & 13 & 84 & 33.55 & 2.58 & 6.4 & 73.7 & 95 \\
    R2 & ccworkflow (opus-5 triage and dispatch) & 3/3 & 15 & 69 & 43.68 & 2.91 & 4.6 & 45.9 & 94 \\
    R3 & ccworkflow (sonnet-5 triage and dispatch, opus-5 integrate, run2) & 3/3 & 16 & 89 & 31.65 & 1.98 & 5.5 & 65.2 & 95 \\ \hline
    R4 & csloop opus-5 & 5/5 & 15 & 43 & 15.69 & 1.05 & 2.9 & 29.9 & 91 \\
    R5 & csloop opus-5 (run2) & 5/5 & 14 & 42 & 13.32 & 0.95 & 3.0 & 25.3 & 91 \\
    R6 & csloop opus-5 (run3) & 4/5 & 12 & 38 & 12.87 & 1.07 & 3.1 & 27.5 & 92 \\
    R7 & csloop sonnet-5 (run2) & 5/5 & 3 & 35 & 5.83 & 1.94 & 11.6 & 130.3 & 91 \\
    R8 & csloop sonnet-5 (run3, incomplete — no agent\_log archived) & 5/5 & 3 & 34 & 6.27 & 2.09 & 11.5 & 148.0 & 91 \\ \hline
    R9 & csloop gpt-5.6-sol (run4) & 5/5 & 4 & 22 & 6.38 & 1.60 & 5.4 & 76.5 & 79 \\
    R10 & csloop gpt-5.6-sol (run5) & 4/5 & 9 & 25 & 9.59 & 1.07 & 2.8 & 42.3 & 82 \\
    R11 & csloop gpt-5.6-sol (run6) & 5/5 & 10 & 38 & 10.06 & 1.01 & 3.8 & 48.8 & 83 \\ \hline
    R12 & ccloop opus-5 & 4/5 & 15 & 69 & 17.76 & 1.18 & 4.6 & 23.5 & 95 \\
    R13 & ccloop opus-5 (run2) & 5/5 & 13 & 82 & 24.50 & 1.88 & 6.3 & 39.7 & 95 \\
    R14 & ccloop sonnet-5 & 1/5 & 3 & 23 & 3.37 & 1.12 & 7.7 & 40.3 & 95 \\
  \bottomrule
\end{tabular}

\end{table*}

\begin{description}
\item[\emph{ccworkflow}] (runs R1--R3) the multi-agent workflow, run by Claude Code. A triage agent selects ready files that do not depend on each other, author subagents translate them in parallel, and a single integrator merges the results and runs the checks. The limits are stated in the prompts. 
\item[\emph{csloop}] (runs R4--R11) the author--reviewer loop of Figure~\ref{fig:csloop}, run by CodeScribe, with at most five cycles. One author agent works through the plan and a reviewer checks its work. The orchestrator enforces the limits and records the author's tool calls for the reviewer. 
\item[\emph{ccloop}] (runs R12--R14) the same author--reviewer loop, with at most five cycles, run by Claude Code. The execution policy is stated in the prompt, the agent has an unrestricted shell, and the author reports its own actions to the reviewer.
\end{description}

The configurations ccworkflow and csloop differ in both orchestration pattern and base agent, so comparing them directly cannot isolate the effect of either factor on cost and performance. The third configuration, ccloop, shares its agent with ccworkflow and its pattern with csloop, so each pairwise comparison varies only one factor.

Comparing ccloop against ccworkflow holds the agent and the model fixed
(Claude Code, Opus~5) and varies only the orchestration pattern. Comparing ccloop
against csloop holds the pattern fixed, and varies only the base tool-calling agent and its execution policy. Together, these two comparisons let us attribute performance differences to specific layers of the agentic stack. Within csloop the orchestrator stays the same while the model changes across different families, so differences there can be attributed to the model and its API format.

Table~\ref{tab:runs} lists the models used in each experiment. The ccworkflow run R2, and the ccloop runs R12 and R13 use Opus~5 in every agent, so they differ only in the orchestration pattern. The csloop runs use three models, Opus~5, Sonnet~5, and GPT~5.6-Sol (the last through an OpenAI-compatible endpoint), to test whether the results carry over to different model tiers and families. The ccloop run R14 repeats the ccloop configuration with Sonnet~5. All CodeScribe loop runs enable reasoning, using adaptive thinking for Opus~5 and Sonnet~5 and effort level high for GPT~5.6-Sol. Claude Code runs use its default effort level of high.

An agentic run is a stochastic process, so a single observation per condition cannot separate an orchestrator or model effect from run-to-run variance. Every configuration was replicated except ccworkflow Opus~5 throughout (R2) and ccloop Sonnet~5 (R14), which are single runs.

\begin{figure*}[h]
  \centering
  \resizebox{0.99\textwidth}{!}{%
  \begin{tikzpicture}
    \node[draw, rounded corners=3pt, fill=black!2, inner sep=8pt] (transcript) {%
      \obeyspaces\ttfamily\footnotesize
      \begin{tabular}{@{}l@{}}
      \color{black}\bfseries{}$\blacktriangleright$ code-scribe loop dev/transformations/mcfm-translate/loop.toml -m <model-name> -v ---reason \\[6pt]%
      \color{black}\bfseries{}$\blacktriangleright$  loop 1 [author] \\%
      \color{black}  $\vdots$   (iters 1--4: read plan/spec, scope initial candidate files) \\[6pt]%
      \color{black}  iter 5 in 1:17.4 \\%
      \color{black!55}    | Let me look at the roadmap metrics to find ready files to translate. I need to see which files are ready (deps=0, \\%
      \color{black!55}    | blind=0, no generated .cpp yet). \\%
      \color{black}    usage  in 1,324  out 334  cache\_write 3,197  cache\_read 9,242  total 14,097 \\%
      \color{black}    $\triangleright$ bash   head -5 dev/tmp/assets/roadmap\_metrics.tsv                    bash exit\_code=0 \\%
      \color{black}    $\triangleright$ bash   grep -v "\textasciicircum{}file" dev/tmp/assets/roadmap\_metrics.tsv | awk -F'...  Error: blocked shell syntax detected \\%
      \color{black}  iter 6 in 6.5s \\%
      \color{black!55}    | Let me use Python to parse the roadmap metrics instead of using awk which is blocked. \\%
      \color{black}    usage  in 1,393  out 218  cache\_write 1,321  cache\_read 12,439  total 15,371 \\%
      \color{black}    $\triangleright$ bash   python3 -c " import csv rows = [] with open('dev/tmp/assets/...  bash exit\_code=0 \\[6pt]%
      \color{black}  $\vdots$ \\[6pt]%
      \color{black}  iter 14 in 1:26.0 \\%
      \color{black!55}    | [ ... ] The key issue is that Fortran's reshape fills arrays in column-major order (leftmost index varies fastest), \\%
      \color{black!55}    | while C++ uses row-major order (rightmost index varies fastest). So when translating the Fortran data initialization \\%
      \color{black!55}    | to C++, I need to account for this difference in how the flattened array maps to the multidimensional indices---element \\%
      \color{black!55}    | 1 in Fortran corresponds to pp(-4,-4,-4,-4), but in C++ I'd need to reverse the index order to get the same logical \\%
      \color{black!55}    | mapping. \\%
      \color{black}    usage  in 6,317  out 2,580  cache\_write 4,713  cache\_read 62,922  total 76,532 \\%
      \color{black}    $\triangleright$ read   software/mcfm/src/Inc/ppmax.f                                 \# path: /home/user/JSS-2026-Example/dev... \\[6pt]%
      \color{black}  $\vdots$   (iters 15--... : author .hpp/.cpp/\_fi.F90, build, update agent\_log.md) \\[6pt]%
      \color{black}  tokens  in 138,516  out 51,409  cache\_write 134,313  cache\_read 2,037,056  total 2,361,294 \\[6pt]%
      \color{black}\bfseries{}$\blacktriangleright$  loop 1 [review] \\%
      \color{black}  $\vdots$ \\%
      \color{black}  tokens  in 28,033  out 6,239  cache\_write 25,279  cache\_read 46,611  total 106,162 \\[6pt]%
      \color{black}  $\vdots$   (loops 2--3 [author]/[review] repeat the same author/review pattern) \\[6pt]%
      \color{black}\bfseries{}$\blacktriangleright$  loop 4 [author] \\%
      \color{black}  $\vdots$ \\%
      \color{black}  tokens  in 156,270  out 89,281  cache\_write 294,227  cache\_read 2,365,838  total 2,905,616 \\[6pt]%
      \color{black}\bfseries{}completed 4/5 loop(s) --- run: 20260724-020711-dd1c55 --- artifacts: .../.codescribe/loop \\%
      \end{tabular}%
    };
  \end{tikzpicture}%
  }
  \caption{An excerpt of a CodeScribe run transcript (transformations/mcfm-translate, captured
    with \file{-v --reason}), showing the loop-cycle and iteration anatomy of
    Figures~\ref{fig:iteration} and~\ref{fig:csloop} as shell output. Reasoning
    summaries are set in the lighter shade used in Figure~\ref{fig:iteration}.
    Vertical dots mark a skipped iteration or phase and an ellipsis a
    truncated path or omitted reasoning text. Nothing between the markers is
    reworded. The run completed 4 of its 5 configured cycles.}
  \label{fig:transcript}
\end{figure*}

\section{Evaluation}
\label{sec:evaluation}

The fourteen runs address the three points raised in Section~\ref{sec:casestudy}, one per subsection below.

Figure~\ref{fig:transcript} shows how an author agent evolves over a csloop run. The prompt cache builds up as the run progresses, so cost grows steadily through a phase. Figure~\ref{fig:evalcombined}f compares the resulting token mix across runs.
The agent also adapts to the shell allowlist. When its \file{awk} pipeline is blocked, it substitutes a Python one-liner in response to the ReAct feedback.
CodeScribe blocks disallowed pipes and metacharacters regardless of which binary issues them, so the agent falls back to more verbose commands.
With Python on the allowlist, agents write reusable tools under \file{dev/tmp}, which can be migrated to \file{dev/tools} after review.
The instruction to write scripts when using Python was a suggestion rather than a constraint, so this behavior was not deterministic.
Each rejection costs an iteration, a tradeoff we discuss in Section~\ref{sec:evalpolicy}.
The transcript also shows the agent reasoning about Fortran and C++ array storage order before writing the array code. The agent eventually abandoned this direction because an \file{FArray} container, which stores C++ arrays in a column-major layout, was already available.
\input{figures/fig_eval}
\begin{table}[t]
  \centering
  \caption{Which top-level \file{src/} module each run's settled files came
    from, counted git-exact. All fourteen runs drew from the same three
    modules (Section~\ref{sec:evaldecision}).
    Table~\ref{tab:coveragemap} maps each run code back to its archived
    directory.}
  \label{tab:coverage}
  \footnotesize
  \setlength{\tabcolsep}{3.2pt}
\begin{tabular}{@{}lrrrr@{}}
  \toprule
  & BDK & Mods & W2jet & Total \\
  \midrule
    R1 & 5 & 3 & 5 & 13 \\
    R2 & 5 & -- & 10 & 15 \\
    R3 & 5 & -- & 11 & 16 \\ \hline
    R4 & -- & 4 & 11 & 15 \\
    R5 & 5 & 4 & 5 & 14 \\
    R6 & -- & 2 & 10 & 12 \\
    R7 & -- & 3 & -- & 3 \\
    R8 & -- & 3 & -- & 3 \\ \hline
    R9 & -- & -- & 4 & 4 \\
    R10 & -- & 4 & 5 & 9 \\
    R11 & 5 & -- & 5 & 10 \\ \hline
    R12 & 5 & -- & 10 & 15 \\
    R13 & -- & 3 & 10 & 13 \\
    R14 & -- & 3 & -- & 3 \\
  \bottomrule
\end{tabular}

\end{table}
\subsection{Cost Metrics and Measurement}
\label{sec:evalresults}
We count files settled from the git branch each run produced, rather than from the checklist the agent kept in \file{agent\_log.md}. The checklist included files edited only to update the build system, whereas reviewing each git branch lets us count converted files exactly.

Costs in Table~\ref{tab:runs} are list prices, not actual charges, since the runs used different billing. 
Per million input / output / cache-write / cache-read tokens, 
Anthropic's Opus~5 rates were \$5 / \$25 / \$6.25 / \$0.50 and Sonnet~5's \$2 / \$10 / \$2.50 / \$0.20. The GPT~5.6-Sol runs used OpenAI's discounted-period rates of \$4 / \$20 / \$5.00 / \$0.40.

The \file{Loops} column of Table~\ref{tab:runs} also differs in kind by orchestrator. Runs under csloop and ccloop end at a fixed author--review budget, an agent completion message, or the approval gate, whereas ccworkflow ends only at the approval gate. \file{Cache} is the cache-read share of input-side tokens, using a common definition across providers.
\subsection{A loop against a multi-agent workflow}
\label{sec:evalloop}
The loop was cheaper than the multi-agent workflow overall. Cost,
tool calls, and tokens per file settled were all lower for most loop runs
(csloop R4--R6, R9--R11, and ccloop R12) than for the workflow runs (R1--R3). The gap is large in cost and token usage and smaller in tool calls (Table~\ref{tab:runs} and Figure~\ref{fig:evalcombined}). Within csloop, an Opus~5 agent kept to a simple pattern, searching with \file{grep} and then reading the lines it found, writing its own Python scripts under \file{dev/tmp} to perform specific tasks not covered by \file{dev/workflow.py}. GPT~5.6-Sol agents explored more instead, relying on \file{read} and \file{glob} to view file contents, resulting in more tool calls. Sonnet~5 agents, by contrast, spent much of their budget on retries, as their bash calls failed more often than in the other two model runs. Agents in ccworkflow relied on \file{cd}/\file{cat} more. Agents in ccloop mostly followed the guidance in the workflow script and behaved like csloop agents, but without architectural enforcement of the orchestration rules their cost remained high compared to csloop.

The R12 (ccloop) and R2 (ccworkflow) runs used the same orchestrator (Claude Code) and model (Opus~5). Both runs settled on the same files. The loop cost about 40\% as much for the same deliverable, in about the same time per file.

The reasons are specific to this task. When many agents write files in
parallel, each one gets its own copy of the same background material, so the
text processed grows with the number of files attempted, while the number
finished depends only on what survives the single integration step. The files in this
task do not depend on each other, but each one is small next to the
material an agent has to read before touching it, so splitting the work
across many agents mostly means reading that material many times over. The
time saved by working in parallel does not compensate for it.

\subsection{A specialized against a general-purpose agent}
\label{sec:evalpolicy}
This comparison holds the agentic approach constant and varies only how the behavioral rules are enforced. csloop (R4--R11) and ccloop (R12--R14) ran similar author--reviewer loops over the same agentic layer. CodeScribe enforced its execution policy architecturally, restricting the number of agent iterations and the context size per author--reviewer loop, checking for repeated tool calls, and protecting files the agent is not allowed to edit. The Claude Code loop gave its agents unrestricted access and stated its rules only through prompted constraints.

As a result, ccloop author phases frequently ran past the bound, one reaching 67 iterations (limit 30), and deviated from the execution policy in their tool use and responses. With no mechanism beyond prompting to enforce the policy, these phases never received the corrective ReAct feedback that would have constrained their behavior. In contrast, csloop saw no phase exceed the bound or deviate from policy.

Claude Code's agent memory management also influenced ccloop's performance. Claude Code saves notes in memory files kept in a user's private configuration and reuses them in later sessions. The memory from R1 helped R12 skip the \file{Mods/} directory and avoid translating code with Fortran idioms, saving tool calls and reasoning effort. It therefore finished faster and at lower cost (Figure~\ref{fig:evalcombined}b). R13 and R14 ran without this memory and processed the files R12 skipped. We discuss the implications in Sections~\ref{sec:evalmodels} and~\ref{sec:evaldecision}. CodeScribe has a similar mechanism, under \file{.codescribe/loop}, for resuming loops, but we ran all CodeScribe cases fresh, so no earlier loop logs influenced the runs.

Agents bounded and checked by the surrounding system were more efficient and stayed within the execution policy, while agents governed only by prompt instructions did not. Reproducible, auditable, and predictably terminating runs therefore require enforced structure rather than prompted constraints.

\subsection{Comparison across models}
\label{sec:evalmodels}
This comparison keeps the orchestrator the same and changes only the model. 
The models differed in their tool-calling behavior, and these differences drove cost. 
The cheapest and most expensive runs in Table~\ref{tab:runs} do not line up with which model tier is nominally cheaper. The tier billed as cheapest produced two very expensive runs per file settled (Sonnet~5, R7 and R8). The pattern is clearer in tool calls. The two Sonnet~5 runs used several times as many tool calls per file as every other csloop run (Table~\ref{tab:runs} and Figure~\ref{fig:evalcombined}(d)), and finished fewer files. This suggests that Sonnet~5 needed many more passes on a file to make the same progress, some of them from rejections and constraints imposed by the bounded shell. The cheapest-billed model was therefore the least efficient per file settled.

The runs from a different family (csloop GPT~5.6-Sol, R9--R11) show that the results carry over to a different provider. R10--R11 performed similarly per file to the Opus~5 runs at a similar token cost, though GPT~5.6-Sol used more tool calls to get there. They also picked work that closely matched what the Opus~5 runs picked (Section~\ref{sec:evaldecision}).

A major difference between the runs originates in a Fortran idiom in the MCFM codebase that has no direct C++ equivalent.
On one file, the runs split on a design decision the agent specification never addressed. The Fortran original declared four named precision kinds.
\begin{lstlisting}[style=mdstyle, title={Mods/types\_mod.f}]
integer, parameter :: sp = selected_real_kind(6)
integer, parameter :: dp = selected_real_kind(15)
integer, parameter :: ex = selected_real_kind(18)
integer, parameter :: qp = selected_real_kind(33)
\end{lstlisting}
Translating these parameters meant giving each kind a real C++ type, including a compiler extension type for extended precision that standard C++ does not define. Two of the three Opus~5 runs made that choice.
\begin{lstlisting}[style=mdstyle, title={Mods/types\_mod.hpp}]
typedef float sp;
typedef double dp;
typedef long double ex;
typedef __float128 qp;
\end{lstlisting}
The third Opus~5 run, and all three GPT~5.6-Sol runs, wrote a plain integer for each kind instead of a real type. One of them named the real type in a comment next to the integer it stored.
\begin{lstlisting}[style=mdstyle, title={Mods/types\_mod.hpp}]
// selected_real_kind(6)  -> 4 byte  real  (float)
const int sp = 4;
// selected_real_kind(15) -> 8 byte  real  (double)
const int dp = 8;
// selected_real_kind(18) -> 10 byte real  (long double)
const int ex = 10;
// selected_real_kind(33) -> 16 byte real  (__float128)
const int qp = 16;
\end{lstlisting}
These integers are the kind values that \file{selected\_real\_kind} returns under gfortran on x86-64. Two GPT~5.6-Sol runs wrote the identical byte counts. The remaining run copied each kind's own precision argument straight through (6, 15, 18, 33), as if it were the value. Either way, the result passed every test. The converted C++ code used the type keywords directly wherever a kind was needed, rather than referencing these constants, so the choice did not affect the benchmark results. However, we count this file as settled when the model made a type-preserving choice, as Opus~5 did above.

Opus~5 and GPT~5.6-Sol runs also differed in two other ways: how much of their input came from cache (Table~\ref{tab:runs}) and how often they ran into the bounded shell's limits. 
The differences in cache efficiency stem from the different API toggles that Anthropic and OpenAI expose for their model families, while the differences in shell-limit encounters reflect each model's own tool-use choices. 
The models had little impact on which files were picked.

\begin{figure*}[t]
  \centering
  \resizebox{0.95\textwidth}{!}{%
\begin{tikzpicture}[
  modMark/.style={draw=evalInkStrong, fill=evalSurface, line width=0.6pt,
    inner sep=0pt},
  modBDK/.style={modMark, circle, minimum size=4.5pt},
  modMods/.style={modMark, rectangle, minimum size=4.5pt},
  modW2jet/.style={modMark, regular polygon, regular polygon sides=3,
    minimum size=6pt},
]
\begin{groupplot}[group style={group size=1 by 1}]
\nextgroupplot[
  width=0.85\textwidth, height=5.7cm,
  axis lines=left, axis line style={draw=evalAxis, line width=0.5pt},
  xmajorgrids, grid style={draw=evalGrid, line width=0.35pt},
  tick label style={font=\footnotesize}, label style={font=\footnotesize, color=evalInk},
  xmin=0, xmax=43.2, ymin=-0.7, ymax=13.700,
  ytick={0,1,2,3,4,5,6,7,8,9,10,11,12,13}, yticklabels={R14,R13,R12,R11,R10,R9,R8,R7,R6,R5,R4,R3,R2,R1},
  y tick label style={font=\scriptsize}, xtick style={draw=none}, ytick style={draw=none},
  xlabel={Elapsed minutes since run start},
  legend columns=3, legend style={at={(0.5,1.03)}, anchor=south, font=\scriptsize,
    draw=none, fill=none, inner sep=1pt, column sep=8pt},
]
\addlegendimage{only marks, mark=*, mark size=2.65pt,
  mark options={draw=evalInkStrong, fill=evalSurface, line width=0.6pt}}
\addlegendimage{only marks, mark=square*, mark size=2.35pt,
  mark options={draw=evalInkStrong, fill=evalSurface, line width=0.6pt}}
\addlegendimage{only marks, mark=triangle*, mark size=3.4pt,
  mark options={draw=evalInkStrong, fill=evalSurface, line width=0.6pt}}
\legend{BDK, Mods, W2jet}
\node[modMods] at (axis cs:7.36,0) {};
\draw[evalAxis, line width=0.5pt] (axis cs:0.654,1) -- (axis cs:9.94,1);
\node[modMods] at (axis cs:0.654,1) {};
\node[modW2jet] at (axis cs:9.94,1) {};
\draw[evalAxis, line width=0.5pt] (axis cs:0.739,2) -- (axis cs:37.9,2);
\node[modW2jet] at (axis cs:0.739,2) {};
\node[modBDK] at (axis cs:37.9,2) {};
\draw[evalAxis, line width=0.5pt] (axis cs:4.46,3) -- (axis cs:20.4,3);
\node[modW2jet] at (axis cs:4.46,3) {};
\node[modBDK] at (axis cs:20.4,3) {};
\draw[evalAxis, line width=0.5pt] (axis cs:0.452,4) -- (axis cs:12.6,4);
\node[modMods] at (axis cs:0.452,4) {};
\node[modW2jet] at (axis cs:12.6,4) {};
\node[modW2jet] at (axis cs:3.28,5) {};
\node[modMods] at (axis cs:1.06,6) {};
\node[modMods] at (axis cs:1.06,7) {};
\draw[evalAxis, line width=0.5pt] (axis cs:0.533,8) -- (axis cs:9.47,8);
\node[modMods] at (axis cs:0.533,8) {};
\node[modW2jet] at (axis cs:9.47,8) {};
\draw[evalAxis, line width=0.5pt] (axis cs:0.622,9) -- (axis cs:10.1,9) -- (axis cs:28.2,9);
\node[modMods] at (axis cs:0.622,9) {};
\node[modW2jet] at (axis cs:10.1,9) {};
\node[modBDK] at (axis cs:28.2,9) {};
\draw[evalAxis, line width=0.5pt] (axis cs:0.683,10) -- (axis cs:11.9,10);
\node[modMods] at (axis cs:0.683,10) {};
\node[modW2jet] at (axis cs:11.9,10) {};
\draw[evalAxis, line width=0.5pt] (axis cs:1.41,11) -- (axis cs:24.1,11);
\node[modW2jet] at (axis cs:1.41,11) {};
\node[modBDK] at (axis cs:24.1,11) {};
\draw[evalAxis, line width=0.5pt] (axis cs:0.811,12) -- (axis cs:23.7,12);
\node[modW2jet] at (axis cs:0.811,12) {};
\node[modBDK] at (axis cs:23.7,12) {};
\draw[evalAxis, line width=0.5pt] (axis cs:0.714,13) -- (axis cs:4.08,13) -- (axis cs:26,13);
\node[modMods] at (axis cs:0.714,13) {};
\node[modBDK] at (axis cs:4.08,13) {};
\node[modW2jet] at (axis cs:26,13) {};
\end{groupplot}
\end{tikzpicture}%
  }
  \caption{What the agent chose to work on and when, across the fourteen
    \file{transformations/mcfm-translate} runs, each collapsed to its selecting
    model (the run's own model for csloop and ccloop, the triage model for
    ccworkflow). Per run, the elapsed time of the first tool call that names
    a file inside a module the run went on to settle in, in module-entry
    order, with marker shape (circle, square, triangle for BDK, Mods, W2jet) marking the module. n/a marks a run that settled nothing.
    Cross-model agreement on the files settled is discussed in
    Section~\ref{sec:evaldecision}.}
  \label{fig:decision}
\end{figure*}
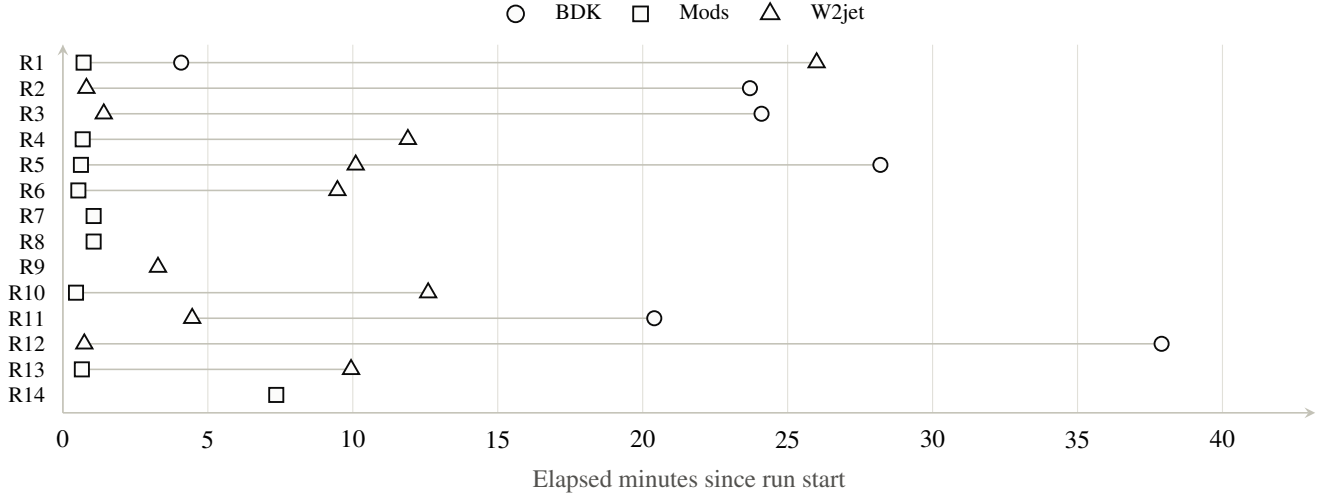

\subsection{Model decision making versus tooling}
\label{sec:evaldecision}

Deterministic tooling under \file{dev/tools} has the agent pick files from a Doxygen map that scores each file with three numbers built from its call graph: how many untranslated files it depends on (deps), whether the parser could analyze it at all (blind), and how many untranslated files call the functions and subroutines in the reference file (fan-in). Fan-in pushes widely-used files to the front of the queue. A file is ready once everything it depends on is translated, and which ready file to work on next is left to the model.

Figure~\ref{fig:decision} shows all fourteen runs picking files in nearly the same order and settling on largely the same files, regardless of which model or orchestrator ran them (Table~\ref{tab:coverage}). At the starting point, only a handful of ready files had a fan-in above one. Almost every other ready file had zero fan-in, so translating it would not make any further file ready.
With so few files worth prioritizing, the convergence of fourteen separate runs on the same ones is expected.

The precision-kinds file discussed above was recommended to every agent with a
fan-in of 8, despite being used in more than 200 files. Doxygen counts this file
only when a kind parameter it defines is used as a function argument, as in
\file{real(nf,dp)}, and not in declarations such as \file{real(dp)::nf},
where the kind almost always appears. Even so, it was the top recommendation and belonged to the \file{Mods/} directory.

What to do with a file like this was also left unclear in the specification, which supported two readings. The file could be translated into a header that compiles, or it could be recognized as not fitting the plan's recipe and skipped as a standing exception. Every CodeScribe run took the first reading. The Claude Code runs, relying on their native memory, took the second. The first Claude Code run flagged the file for human review and wrote a no-op for it in memory. Citing that flag, every later Claude Code run skipped the file and the \file{Mods/} directory entirely, moving on to other directories (Figure~\ref{fig:decision}). A set of csloop runs with Kimi-K3, archived under \file{evals/experiments} \citep{jss_example_2026}, also skipped the precision-kinds file. Unlike the Claude Code runs (R1--R3, R12), however, they went on to work on the other files in \file{Mods/}. R13 and R14, run after purging the Claude Code memory, processed both \file{Mods/} and the precision-kinds file, as the Opus~5 and GPT~5.6-Sol csloop runs did.

\section{Conclusion}
\label{sec:conclusion}

Scientific software is expensive to maintain and sensitive to error, and AI agents can already carry out substantial autonomous work on it.
Using that capability without eroding domain understanding and correctness guarantees requires structure around the agent.
We designed such a collaborative orchestration structure and compared its variants on a single task, the translation of MCFM.

In our approach, humans retain ownership of intent and acceptance, and numerical correctness is specified and enforced at an explicit verification boundary. Hallucinated progress is contained by checks that the workflow must confirm before a file counts as verified. Five design principles connect the orchestration architecture with the agentic layer to run a multi-agent workflow or a two-agent loop.

We ran fourteen versions of the same translation task, changing the orchestrator (CodeScribe vs Claude Code) and the models (Opus~5, Sonnet~5, GPT~5.6-Sol), to identify which parts of the design determined the outcome. A single author--reviewer loop matched the output of the multi-agent workflow at roughly one third of the cost per file settled. Multiple parallel agents pay off only when the work and its context are independent (Section~\ref{sec:evalloop}).

The models we tested chose similar files, largely because the ranking tool used by the orchestrator determined which files were ready to translate. Tool usage was enforced through the agent execution policy (CodeScribe) or through prompt-based instructions (Claude Code), which influenced the outcome. One result was driven by an orchestrator memory note that influenced agent decision-making in three later runs without leaving a trace in our records. The oversight was ours, but a mechanism that can introduce a correct assumption across runs can introduce an incorrect one the same way.

Because the specification, plan, and tools drive the workflows, a mistake in that layer affects the outcome of every run. A gap in the shared specification left the C++ representation of the Fortran kind parameters (sp, dp, ex, qp) unresolved, and two incompatible answers: a real extended-precision type, and a placeholder integer were produced by the models. Both passed the full benchmark suite, since the converted code never referenced these constants directly. Only the first Claude Code run caught the gap, by flagging the file for human review rather than resolving it itself (Section~\ref{sec:evaldecision}). More generally, a file that does not fit the agentic specifications should be left entirely to human review.

Shared agentic layers can help teams organize their science campaigns and promote systematic use of AI to improve productivity. 
The author-reviewer loop can be applied to deploy simulation campaigns, run analyses, and perform work in manageable portion, which lets humans supervise AI work effectively. 
Several aspects of agentic layers remain open and may continue to shape their design. 
One is the format of the model API. 
The model API exposes controls for prompt caching and tool use, which account for the cache efficiency of CodeScribe (Section~\ref{sec:evalmodels}). 
Reasoning summaries, also available through the API, allow direct inspection of model reasoning and support human-AI collaboration. 
Exposing reasoning through an API introduces security risks, and recent work has proposed mitigations~\citep{panfilov2026stealing}. 
Such work will likely influence how open-source LLMs are used and may lead to a common model API for scientific applications.

Agent execution policy is a further open problem. A recent report found that multi-agent systems operating at scale circumvented well-enforced guardrails to complete the work their specifications prescribed \citep{metr2026huggingface}. Specification and policy design therefore require care. Treating this design as a collaborative process allows agentic systems to be evaluated as scientific instruments prior to autonomous deployment.

\section*{Acknowledgements}
\label{sec:acknowledgements}
We thank John Campbell and Stefan Höche for many fruitful discussions on MCFM and its design, along with broader conversations on higher-order QCD calculations for hadron colliders. This material is based upon work supported by the U.S. Department of
Energy, Office of Science, Office of Advanced Scientific Computing
Research, Scientific Discovery through Advanced Computing (SciDAC) program
under contract DE-AC02-07CH11359. Max Knobbe and Pedro Machado are supported by Fermi Forward Discovery Group, LLC under Contract No. 89243024CSC000002 with the U.S. Department of Energy, Office of Science, Office of High Energy Physics. Akash Dhruv is supported by UChicago, LLC, under Contract No. DE-AC02-06CH11357 with the U.S. Department of Energy, Office of Science.

\appendix
\section{Dataset Archive}

The evaluations in Section~\ref{sec:evaluation} were carried out using the Lab Notebook archived on Zenodo \citep{jss_example_2026}. The archive contains transformations and datasets for agent trajectories described in Section~\ref{sec:doe}. These trajectories are organized by date, with separate agent logs for the CodeScribe and Claude Code runs.

The two orchestrators record their agent logs in different formats. Claude Code supplies \file{jsonl} logs, which can be parsed with Python scripts to extract token usage and tool calls for each assistant message; each such message is treated as one agent iteration. CodeScribe's loop agents store their logs in \file{toml} files containing detailed records of tool-call events, token usage, and run information. These logs are collected by a metadata agent, which is triggered after a run is complete. The metadata agent copies the run logs to the \file{evals/experiments} directory and creates a corresponding GitHub branch that records the MCFM code changes in its repository. For all agents, the reasoning summaries and hashes returned by Anthropic's Messages API have been scrubbed from the logs.

The metadata agent uses \file{evals/archive.toml}, which provides simple instructions on how to discover, archive, and label the run logs. It is supplied as a task file to a CodeScribe agent:
{\small
\begin{verbatim}
code-scribe agent archive.toml -m <model>
\end{verbatim}
}

The dataset is released under the MIT license, and a dataset card summarizing its contents is included in the Zenodo archive. A set of scripts under \file{evals/analysis} reproduce the study in this paper, and the dataset can also be used to carry out further evaluations. We encourage users to follow the policies of the respective model providers when using and evaluating the dataset.

The experiments are reproducible. The reference Lab Notebook does not contain the converted MCFM code itself, but it does include scripts that clone the corresponding GitHub repository under \file{software/mcfm}.

\section{Declaration of Generative AI Use}

In the course of this work, we used artificial intelligence tools to support both the research process and the analysis of data. This included developing workflow scripts, exploring agentic AI approaches, and assisting with data cleaning and statistical analysis. We reviewed and verified all AI-generated outputs and take full responsibility for the accuracy of the results and the integrity of the work.

\bibliographystyle{elsarticle-harv}
\bibliography{references}

\end{document}